\documentclass[preprint,pra,nofootinbib]{revtex4}

\newcommand{\be}{\begin{equation}}
\newcommand{\ee}{\end{equation}}
\newcommand{\ba}{\begin{eqnarray}}
\newcommand{\ea}{\end{eqnarray}}
\newcommand{\br}{{\bf r}}
\newcommand{\bR}{{\bf R}}
\newcommand{\li}{Li^{3+}}
\newcommand{\pb}{\bar{p}}
\usepackage{graphicx}% Include figure files
\usepackage{dcolumn}% Align table columns on decimal point
\usepackage{bm}% bold math

\begin{document}

%\preprint{APS/123-QED}

\title{Search for Long-lived States in Antiprotonic Lithium}% Force line breaks with \\

\author{J. R\'{e}vai}

\affiliation{%
Research Institute for Particle and Nuclear Physics\\
H-1525 Budapest, P.O.B. 49, Hungary
}%

\author{V. B. Belyaev}
\affiliation{
Bogolyubov Laboratory of Theoretical Physics\\
Joint Institute for Nuclear Research\\
141980 Dubna, Russia% with \\
}%

\date{\today}% It is always \today, today,

\begin{abstract}
The spectrum of the ($\li+\pb+2e$) four-body system was calculated
in an adiabatic approach. The two-electron energies were
approximated by a sum of two single-electron effective charge
two-center energies as suggested in \cite{Briggs}. While the
structure of the spectrum does not exclude the existence of
long-lived states, their experimental observability is still to be
clarified.
\end{abstract}

\maketitle

\section{\label{sec:level1}Introduction}

One of the most impressive success-stories of the last decade in
few-body physics are the high-precision experimental and
theoretical studies of long-lived states in antiprotonic helium
\cite{exp,kor,elan,kam}. In view of this fact it is natural to
pose the question, whether such long-lived antiprotonic states can
exist in other systems, too. There were some experimental attempts
to observe delayed components in annihilation products after
stopping of slow antiprotons in different media \cite{wid},
however, except for helium, no evidence of long-lived states was
found. Theoretical predictions concerning the possible existence
of such states could largely increase the willingness of
experimentalists to go on with their attempts.

There are two possible directions in which further candidates for
long-lived antiprotonic states could be searched for. First, one
could consider atoms with electron structure similar to helium,
that is, noble gases with closed outermost electron shells. This
possibility was examined in some detail in \cite{Briggs} with an
essentially negative answer concerning the possibility of
formation of long-lived states after antiproton capture in noble
gases.

The second possibility is to consider the next simplest atom, the
lithium, which has three electrons, one of which could be replaced
by the antiproton. In order to get an idea of the possibility of
the occurrence of long-lived states in this four-body system, we
have performed a semi-quantitative calculation of its level
structure.

\section{Calculation Method}

The Hamiltonian of the $(\li+\pb+2e)$ four-body system can be
written as
 \be
 \hat{H}=-{1\over {2M}}\Delta_{\bR}-{3\over R} +
 \hat{h}^{(2)}(\br_1,\br_2;\bR),
 \ee
 with the two-electron Hamiltonian
 \be
 \hat{h}^{(2)}(\br_1,\br_2;\bR)=\hat{h}^{(1)}(\br_1;\bR) + \hat{h}^{(1)}(\br_2;\bR) +
 {1\over|\br_1-\br_2|}
 \ee

The single-electron two-center Hamiltonian $\hat{h}^{(1)}$
corresponds to the electron motion in the field of $\li$ and $\pb$
fixed at a distance $\bR$:
 \be
 \hat{h}^{(1)}(\br_n;\bR)=-{1\over{2}}\Delta_{\br_i}-{3\over
 |\br_i-\alpha\bR|} + {1\over|\br_i+(1-\alpha)\bR|}
 \ee
 In eqs.(1-3) \bR\  is the vector pointing from $\li$ to $\pb$,
 while the $\br_i$ are the electron coordinates measured from the
 $\li-\pb$ center of mass.
 \footnote{ These are not exactly the Jacobian coordinates of
 the system, however, the extra terms in the Hamiltonian arising from the difference
 are very small and the accuracy of our calculation does not
 necessitate their consideration}
  $M$ is the reduced mass of $\li$ and
 $\pb$:
 \[ {1\over M}= {1\over m_{\li}} + {1\over m_{\pb}},\]
  while $\alpha$ is defined as
  \[ \alpha={m_{\pb}\over m_{\li} + m_{\pb}}\ \
  \footnote{All masses are expressed in units of electron
mass}
  \]

  To calculate the spectrum of this four-body system or, at
least, a part of it, we have used a Born-Oppenheimer-like (BO)
approximation, in which the solution of the Schr\"{o}dinger-equation
 \be
  (\hat{H} - E^J_n)\Psi^J_n(\br_1,\br_2,\bR)=0
 \ee
is attempted in two successive steps. First, the equation
 \be
 (\hat{h}^{(2)}(\br_1,\br_2;\bR) -
 \varepsilon_{n\mu}^{(2)}(R))\Phi_{n\mu}(\br_1,\br_2;\bR) = 0 ,
 \ee
 describing the motion of two electrons in the field of $\li$ and
 $\pb$ separated by a fixed vector \bR, has to be solved. The
 solutions $\Phi_{n\mu}$ are characterized by the conserved
 quantum number $\mu$ -- the sum of the electron angular momentum
 projections on the \bR\  direction. Next, the total wave function
 $\Psi^J_n(\br_1,\br_2,\bR)$ is approximated as a single product:
  \be
  \Psi^J_n(\br_1,\br_2,\bR)={u^J_{n\mu}(R)\over R}
  D^J_{-M_J,-\mu}(\phi,\Theta,0)\Phi_{n\mu}(\br_1,\br_2;\bR),
  \ee
  where the Wigner's $D$-functions are needed to ensure correct
  angular momentum quantum numbers $JM_J$ for the total wave
  function. The energy eigenvalues $E^J_{n\mu}$ are calculated from
  the radial equation
  \begin{widetext}
   \be
   \left(-{1\over 2M}{d^2\over dR^2} + {J(J+1)-2\mu^2\over 2MR^2} -
   {3\over R} +\varepsilon_{n\mu}^{(2)}(R) -
   E^J_{n\mu}\right)u^J_{n\mu}(R)=0,
   \ee
   \end{widetext}
 which is obtained by substituting Eq.(6) into Eq.(4), multiplying by
 $D^J_{-M_J,-\mu}(\phi,\Theta,0)^*\Phi_{n\mu}(\br_1,\br_2;\bR)^*$
 and integrating over the electron coordinates and the angular
 variables $(\phi,\Theta)$ of \bR. In deriving Eq.(7), according
 to the adiabatic approximation, terms containing the derivatives
 of $\Phi_{n\mu}(\br_1,\br_2;\bR)$ with respect to \bR\  are
 neglected. In BO type calculations usually the lowermost electron
 configurations are used, for which $\mu=0$ (so called
 $\sigma$-term) in which case the $D$-function in Eq.(6) reduces
 to a spherical harmonics $Y_{JM_J}$.

 For a one-electron problem Eq.(5) is replaced by
 \be
(\hat{h}^{(1)}(\br_i;\bR) -
 \varepsilon_{n\mu}^{(1)}(R))\varphi_{n\mu}(\br_i;\bR) = 0
 \ee
 and it can be solved exactly by separation of variables in
 spheroidal coordinates $\varphi_{n\mu}(\br_i;\bR)$ being the
 well-known two-center wave functions. In our case, however, the
 problem of two interacting electrons in the field of two fixed
 Coulomb-centers is not solvable exactly and the determination of
 eigenvalues and eigenfunctions necessitates a highly non-trivial
 calculation. One could think of a variational approach of the
 type \cite{ahlr}; another possibility could be the diagonalization
 of the electron-electron interaction on the basis of two-center
 functions. In both cases the calculations are quite cumbersome
 and slowly converging when the size of the basis is increased.
 Since we believe, that the question of possible existence of
 metastability in antiprotonic lithium can be studied by a
 semi-qualitative exploration of its spectrum, we have chosen a
 simpler, elegant and efficient method of calculation for $\varepsilon_{n\mu}^{(2)}(R)$
 proposed in~\cite{Briggs}.

 The main idea of the method is to represent the two-electron
 energy eigenvalue as a sum of two single-electron energies:

 \begin{equation}
 \varepsilon^{(2)}(R)=\varepsilon^{(1)}_{Z_1,Z_2}(R) +
 \varepsilon^{(1)}_{z_1,z_2}(R),
 \footnote{Here and in the following, where no
 confusion can arise, the indices $(n\mu)$ are omitted}
 \end{equation}

 where $\varepsilon^{(1)}_{Z_1,Z_2}(R)$ and $
\varepsilon^{(1)}_{z_1,z_2}(R)$  are energy eigenvalues of
two-center equations of the type (8) with effective charges
$(Z_1,Z_2)$ and $(z_1,z_2)$ instead of the physical charges
$(3,-1)$. The effective charges $(z_1,z_2)$ are chosen in such a
way, that $\varepsilon^{(1)}_{z_1,z_2}(R=0)$ and
$\varepsilon^{(1)}_{z_1,z_2}(R=\infty)$ should reproduce the
experimental values of the first ionization potentials of $He$
atom and $Li^+$ ion, respectively. Thus we get $z_1=2.3578$ and
$z_2=-1.0135$. As for $(Z_1,Z_2)$, the corresponding two-center
eigenvalues in the $R=0$ and $R=\infty$ limits should reproduce
the second ionization potentials of $He$ and $Li^+$, and the
second electron is in this case the last one, therefore the
physical values $Z_1=3$ and $Z_2=-1$ were taken. The
electron-electron repulsion is taken into account in this method
by the deviation of the effective charges $(z_1,z_2)$ from their
physical (integer) values. In this way the approximate
$\varepsilon^{(2)}(R)$ of Eq.(9) reproduces the experimental
two-electron binding for the two limiting cases $R=0$ and
$R=\infty$, while for intermediate $R$-values the solution of the
corresponding (effective) two-center problems seems to provide a
reasonable interpolation prescription. This approach has been
checked in the case of $(He^{2+}+\pb+2e)$ and $(\li+\pb+2e)$
systems; its results were compared with those of a detailed
variational calculation of \cite{ahlr}. Results of the
calculations are presented in Fig.(1). The agreement of
$\varepsilon^{(2)}(R)$ obtained from Eq. (9) with the variational
values is amazingly good in a wide range of $R$. The same
procedure was applied to calculate the energy of the first excited
electron configuration, where the limiting cases were adjusted to
reproduce the energies of the first excited $(1s2s)$ states of
$He$ and $Li^+$.
\begin{figure*}
\includegraphics[scale=0.6,angle=-90]{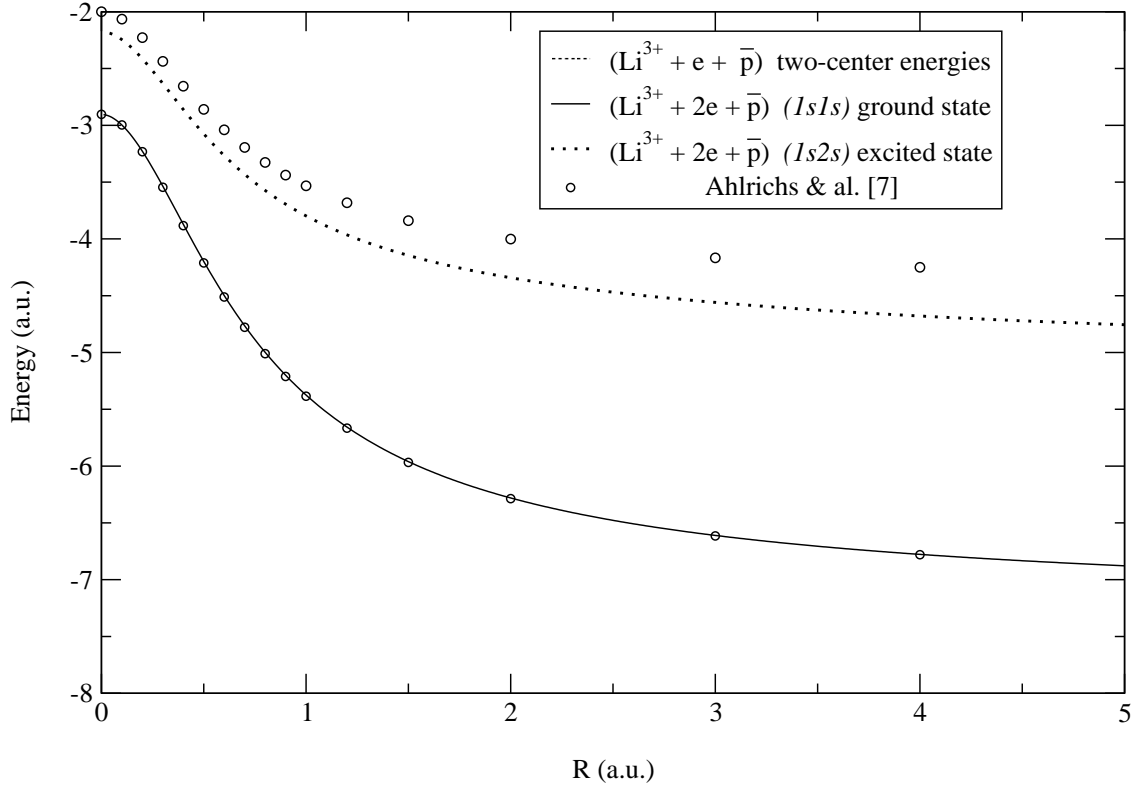}
\caption{\label{fig1} Electronic energies of the ($\li+\pb+2e$)
and ($\li+\pb+e$) systems}
\end{figure*}

Having obtained the electronic energies $\varepsilon^{(2)}(R)$ the
effective potentials of Eq.(7)
 \be
 v^{Jn}_{\text{eff}}(R)={J(J+1)\over 2MR^2}-{3\over
 R}+\varepsilon^{(2)}_n(R) ,
 \ee
can be calculated. Here the index $\mu=0$ was omitted, while the
electron configuration label $n$ can take the values
$n=(1s1s),(1s2s)$. For some values of the total angular momentum
$J$ the effective potentials are shown in Fig.(2) for the ground-
and first excited electron configuration.
\begin{figure*}
\includegraphics[scale=0.6,angle=-90]{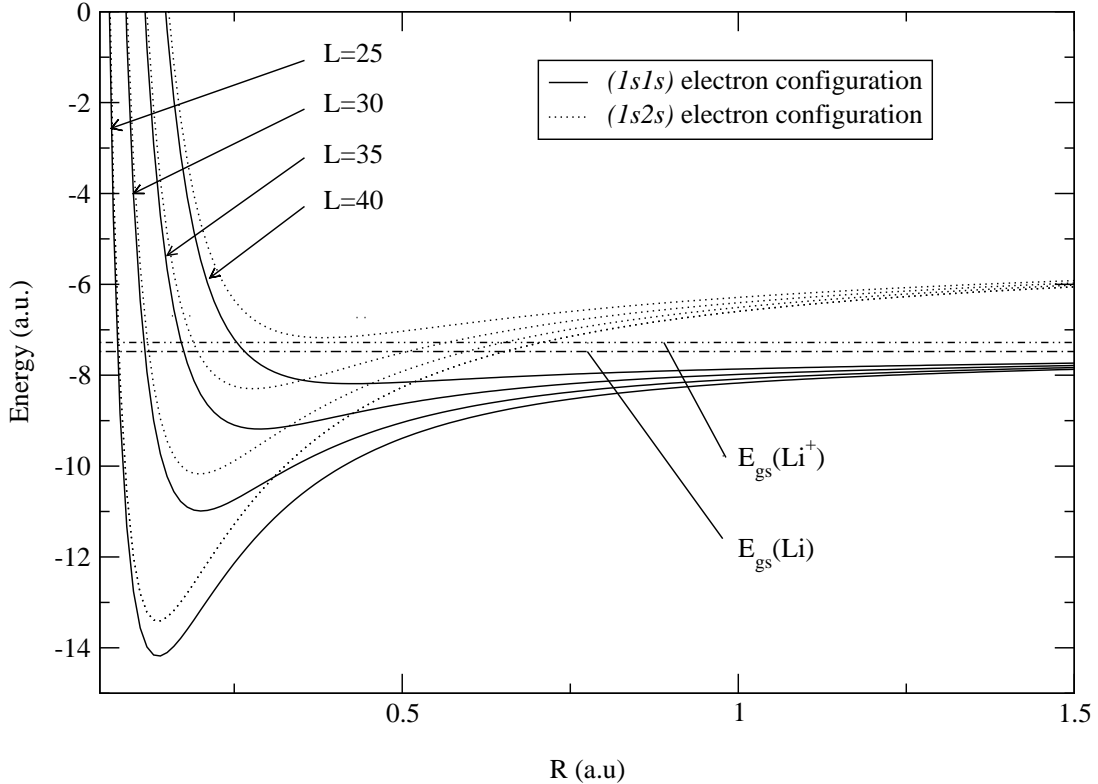}
\caption{\label{fig2} Effective potentials in the ($\li+\pb+2e$)
system }
\end{figure*}
The energy eigenvalues $E^{J\nu}_n$ are then calculated by solving
Eq.(7) with these effective potentials, and the "vibrational"
quantum number $\nu$ is introduced to distinguish among the states
with the same $J$-value.
\section{Results and Discussion}
The resulting spectrum of the $(\li+\pb+2e)$ system is shown in
Fig.(3). Apart from the energy levels of the initial system (full
black circles) we have shown also the energies of daughter states
which can be formed after Auger-emission of one or two electrons.
The open circles correspond to the energies of the $(\li+\pb+e)$
three-particle system, while the open squares are the
hydrogen-like two-particle energies of the $(\li+\pb)$ system.
\begin{figure*}
\includegraphics[scale=0.9]{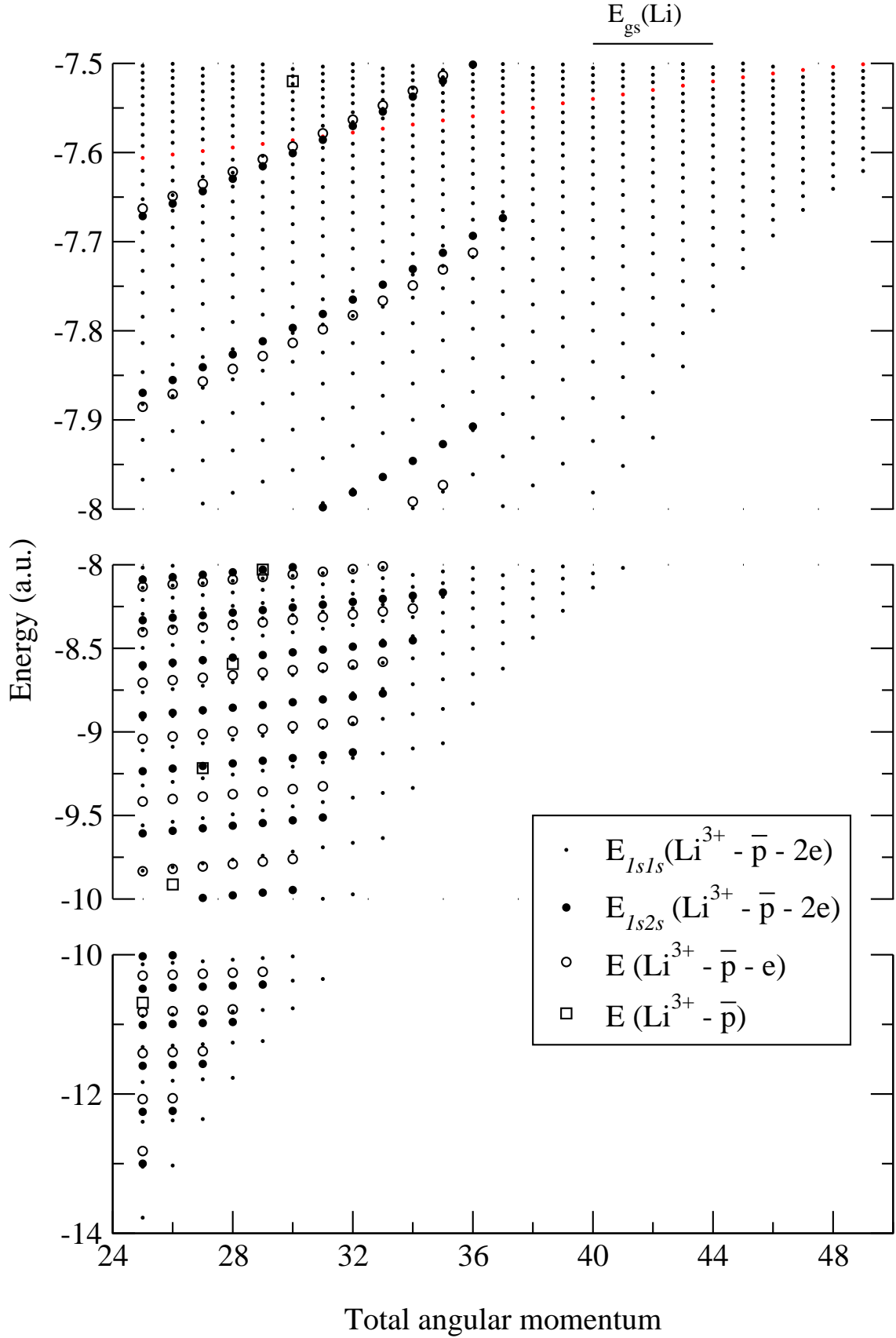}
\caption{\label{fig3} Spectra of antiprotonic lithium atoms}
\end{figure*}
Comparing the spectra of Fig.(3) with the well-known spectra of
the $He$ atomcules (see e.g. \cite{shim}), we can find some
apparent similarities and differences. The basic similarity can be
formulated as follows: there are many states in the spectrum, from
which Auger-emission is possible only with large electron orbital
momentum and therefore is strongly suppressed. This could be one
reason for metastability of these states; of course, this is only
a necessary condition and by no means a sufficient one.

The basic difference, on the other hand, is the much higher
density of states in the expected capture region (around the $Li$
atom ground state energy) which is due to the essential difference
in the electron structure of $He$ and $Li$: the last electron is
strongly bound in $He$, while very loosely in $Li$.

It can be noted, that the spectrum of the $(\li+\pb+e)$ system
(open circles) strongly resembles the $He$ atomcule spectrum,
therefore long-lived states in an isolated $(\li+\pb+e)$ system
could be certainly expected. However, in contrast to the $He$
case, this system is charged and thus its interaction with atoms
of the surrounding medium might be more violent, leading to a
faster collisional de-excitation of these states.
\section{Conclusions}
We have calculated the spectra of ($\li+\pb+2e$) and ($\li+\pb+e$)
four- and three-body systems. Although the structure of the
obtained spectra allows the existence of long-lived states, our
calculations do not put us in a position to make definite
statements about their experimental observability. This latter
depends on several further factors, as well. One is the formation
mechanism: for the time being we have no information about the
population rate of the huge amount of states in the vicinity of
expected antiproton capture energy. The physics of formation of
the $(\li+\pb+2e)$ system in the reaction
\[ (\li+3e)\ + \pb\ \longrightarrow (\li+\pb+2e)\ +\ e \]
is quite different from the analogous process in $He$ due to the
large difference in the binding energies of the outermost
electron. In the $Li$ atom the first ionization potential is only
$0.198\ a.u.$, which means that even adiabatic ionization is
possible: when the distance between the antiproton and $Li$ atom
becomes less than $5\ a.u.$, the binding energy of the last
electron in their common field becomes zero and the electron is
"pushed" into a continuum state.

Another unknown factor is the way, how the eventually formed
$(\li+\pb+2e)$ systems interact with the media atoms and what is
the role of collisional de-excitation in their life-time. In any
case, in order to reduce the undesired effect of this factor,
probably, the experiments looking for long-lived states should be
performed in dilute vapors of $Li$.

\begin{acknowledgments}
One of the authors (VB) wishes to thank the NATO senior fellowship
grant {\bf 1004/NATO/01}, which made possible his visit to
Hungary, while the other author (JR) is grateful for the OTKA
grants T 026244 and T 029440.
\end{acknowledgments}

%\newpage %Just because of unusual number of tables stacked at end

\begin{references}
\bibitem{exp} T.~Yamazaki, N.~Morita, R.~S.~Hayano, E.~Widmann, and
J.~Eades, Physics Reports {\bf 366}, 183(2002)
\bibitem{kor} V.~I.~Korobov, Phys. Rev. A {\bf 54},R1749(1996);\\
V.~I.~Korobov, D.~Bakalov, and H.~J.~Monkhorst, \textit{ibid} {\bf
59},R919(1999)
\bibitem{elan} N.~Elander, and E.~Yarevsky, Phys. Rev. A {\bf
56},1855(1997); \textit{ibid}~{\bf 57},2256(1998)
\bibitem{kam} N.~Yamanaka, Y.~Kino, H.~Kudo, and M.~Kamimura,
Phys. Rev. A {\bf 63},012518(2001)
\bibitem{wid} E.~Widmann \textit{et al}, Phys. Rev. A {\bf
51},2870(1995)
\bibitem{Briggs} J.~S.~Briggs, P.~T.~Greenland, and
E.~A.~Solov'ev, J. of Phys. B {\bf 32},197(1999)
\bibitem{ahlr} R.~Ahlrichs, O.~Dumbrais, H.~Pilkuhn, and
H.~G.~Schlaile, Z. Phys. A {\bf 306},297(1982)
\bibitem{shim} I.~Shimamura, Phys.Rev. A {\bf 46},3776(1992)
\end{references}

\end{document}